\documentclass{PoS}

\usepackage{xspace}
\usepackage{siunitx}
\usepackage[export]{adjustbox}

\newcommand{\psrls}{PSR B1259--63/LS 2883\xspace}
\newcommand{\hess}{H.E.S.S.\xspace}

\DeclareSIUnit\year{yr}
\DeclareSIUnit\erg{erg}
\DeclareSIUnit\au{AU}
\DeclareSIUnit{\msun}{\mbox{$M_{\odot}$}}
\DeclareSIUnit{\rsun}{\mbox{$R_{\odot}$}}
\DeclareSIUnit\ev{eV}
\DeclareSIUnit\kev{\kilo\ev}
\DeclareSIUnit\gev{\giga\ev}
\DeclareSIUnit\tev{\tera\ev}

\title{\hess II observations of the 2014 periastron passage of \psrls}

\ShortTitle{\psrls in 2014 with \hess II}

\author{\speaker{C. Romoli} $^a$, P. Bordas$^b$, C. Mariaud$^c$ and T. Murach$^d$ for the \hess Collaboration\\
\llap{$^a$} Dublin Institute for Advanced Studies, 31 Fitzwilliam Place, Dublin 2, Ireland \\
\llap{$^b$} Max-Planck-Institut f\"ur Kernphysik, P.O. Box 103980, D 69029 Heidelberg, Germany \\
\llap{$^c$} Laboratoire Leprince-Ringuet, Ecole Polytechnique, CNRS/IN2P3, F-91128 Palaiseau, France \\
\llap{$^d$} DESY, D-15738 Zeuthen, Germany\\

E-mail: \email{romolic@cp.dias.ie}}

\abstract{\psrls is a gamma-ray binary system composed of an O9.5Ve main sequence star, LS 2883, and a \SI{47.8}{\milli\second} spinning neutron star in a highly eccentric \SI{3.4}{\year} orbit (eccentricity $e = 0.87$). \psrls is so far the only gamma-ray binary in which the compact object has been firmly identified. \hess observed this system around its periastron passages in 2004, 2007, 2011 and 2014. For this latter event, a detailed campaign was organised making use of the new capabilities of \hess II, in particular its improved sensitivity and a lower energy threshold. This campaign covered for the first time the time of periastron and parts of the orbit so far unexplored at VHE energies, and included as well observations during the GeV flare observed contemporaneously with the Fermi-LAT. The analysis of the \hess II data indicates a relatively high TeV flux during this GeV flare and also at orbital phases preceding the first neutron star crossing of the circumstellar disk. These results will be summarised and discussed in the context of previous models attempting to explain the complex gamma-ray emission from this source.}

\FullConference{35th International Cosmic Ray Conference --- ICRC2017\\
		10--20 July, 2017\\
		Bexco, Busan, Korea}

\begin{document}

\section{Introduction}

The gamma-ray binary \psrls is a system comprising a pulsar in orbit around the O9.5Ve star LS~2883. One of the peculiarities of this source is that it is the only member of the small class of gamma-ray binaries for which the nature of the compact object has been established. The pulsar has a rotational period of \SI{48}{\milli\second} with an associated spin-down luminosity $\dot{E} \approx \SI{8e35}{\erg\per\second}$ \cite{1992ApJ...387L..37J,2013A&ARv..21...64D}. The companion star LS~2883 has a mass between 20 and 30 M$_{\odot}$ and drives a stellar wind and an equatorial outflow which forms a disk around the star which extends to at least 10 stellar radii \cite{1992ApJ...387L..37J,2011ApJ...732L..11N,2014MNRAS.439..432C}.

The orbital period of the system of \SI{3.4}{\year} is one of the largest in the gamma-ray binaries sample detected so far. The orbit itself is very eccentric with an eccentricity parameter $e = 0.87$, with an orbital separation is 13.4 Astronomical Units (\si{\au}) at apastron and less than \SI{1}{\au} at periastron \cite{1992ApJ...387L..37J,1998MNRAS.298..997W,2004MNRAS.351..599W}.

Because the plane of the orbit is inclined with respect to the plane of the circumstellar disk, the pulsar crosses the disk twice, shortly before and after periastron. Its interaction with the stellar wind and the circumstellar disk gives rise to an enhancement of the non-thermal emission that is visible across the full electromagnetic spectrum, from radio to gamma rays (see e.g. Fig.~1 in \cite{2014MNRAS.439..432C}). The times of the crossing with respect time of periastron ($t_{\rm per}$) are inferred from the time when the pulsed signal of the neutron star either vanishes or reappears. These times correspond to $t_{\rm per}-\SI{15}{\day}$ and $t_{\rm per}+\SI{15}{\day}$ \cite{1996MNRAS.279.1026J}. In correspondence to these times there is a clear brightening of the X-ray and radio emission so that a double-peak structured light curve is formed \cite{2014MNRAS.439..432C}.

At \si{\gev} energies the source was detected by the \textit{Fermi}-LAT for the first time during the periastron passage of 2010/2011. In this energy band a strong, unexpected enhancement of the emission starting approximately \SI{30}{\day} after periastron was detected. This flare lasted for more than one month. At the peak, the emitted power in gamma rays almost matched the total spin-down luminosity \cite{2011ApJ...736L..10T,2011ApJ...736L..11A}. This flaring event was detected again around the periastron passage in 2014 \cite{2015MNRAS.454.1358C} strengthening the hypothesis of a periodic phenomenon. However, the explanation of this phenomenon is still unclear, with various possible explanations brought forward \cite{2015MNRAS.454.1358C,2011ApJ...742...98K,2012ApJ...752L..17K}.

At very high energies, the \hess Collaboration discovered the source during the periastron passage in 2004 and recorded also the subsequent passages in 2007, 2010/2011 and 2014. Any observation at other orbital phases did not reveal a detectable signal \cite{2005A&A...442....1A,2009A&A...507..389A,2013A&A...551A..94H,2015arXiv150903090R,2016arXiv161003264B}.

Due to the visibility constraints of ground-based telescopes, during each observation campaign it was possible to probe only parts of the orbit around each periastron. A double-peak profile of the \si{\tev} light curve emerges when merging the data from 2004, 2007 and 2011 into a single phase-folded light curve, similar to what is seen at X-ray energies. In both the \si{\tev} and X-ray light curves a local flux minimum close to $\sim t_{\rm per}$ is observed \cite{2013A&A...551A..94H}. However, observations were still missing at the exact time of the periastron passage and also simultaneous observations together with experiments detecting the \si{\gev} flare were yet to be performed\footnote{The 2004 campaign covered that orbital phase, but at the time there was no \si{\gev} telescope available.}.

After several improvements of the analysis software, in addition to the results reported already in \cite{2015arXiv150903090R,2016arXiv161003264B}, an updated analysis of all the \hess campaigns on this object is presented, showing for the first time a phase-folded lightcurve of \psrls for energies above \SI{1}{\tev} using all available data obtained with the \hess-I telescopes.

\section{\hess Observations in 2014}

The \hess experiment is a hybrid array of Imaging Air Cherenkov Telescopes (IACTs) situated in the Khomas Highland in Namibia at \SI{1800}{\meter} above sea level. Until 2012 it consisted of four \SI{12}{\meter} telescopes (CT1--4) arranged in a square with a side length of \SI{120}{\meter}. In 2012 an additional \SI{28}{\meter} telescope (CT5) was added to the centre of the array with the aim of lowering the energy threshold of the experiment. The data taken during the 2014 observation campaign targeted on \psrls were analysed in different ways depending on the telescopes that were involved. The terminology used in the following paragraphs is:
\begin{itemize}
  \item CT1--4 STEREO analysis --- uses only data obtained with the \SI{12}{\meter} telescopes. Even if the CT5 was participating in the observation, this information is ignored.
  \item CT1--5 STEREO analysis --- uses the information from CT5 plus any one (or more) of the \SI{12}{\meter} telescopes.
  \item CT5 MONO analysis --- uses only the information obtained with CT5.
\end{itemize}

In these proceedings we are showing the results from \SI{62.2}{\hour} using only CT5 MONO data, \SI{63.3}{\hour} of CT1--5 STEREO observations and \SI{68.1}{\hour} of CT1--4 STEREO observations from 2014, and additionally the results from reanalyses of the archival data from previous observation campaigns. The results reported here were obtained from analyses performed with the ImPACT reconstruction \cite{2014APh....56...26P,2015ICRC...34..826P}.

\begin{figure}
  \centering
  \includegraphics[width=0.45\textwidth,valign=t]{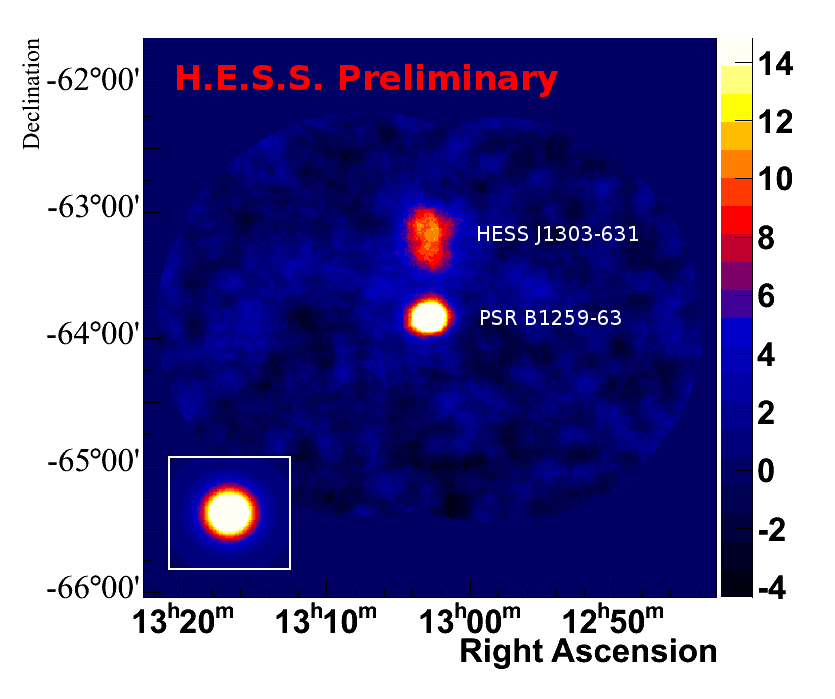}\qquad
  \includegraphics[width=0.45\textwidth,valign=t]{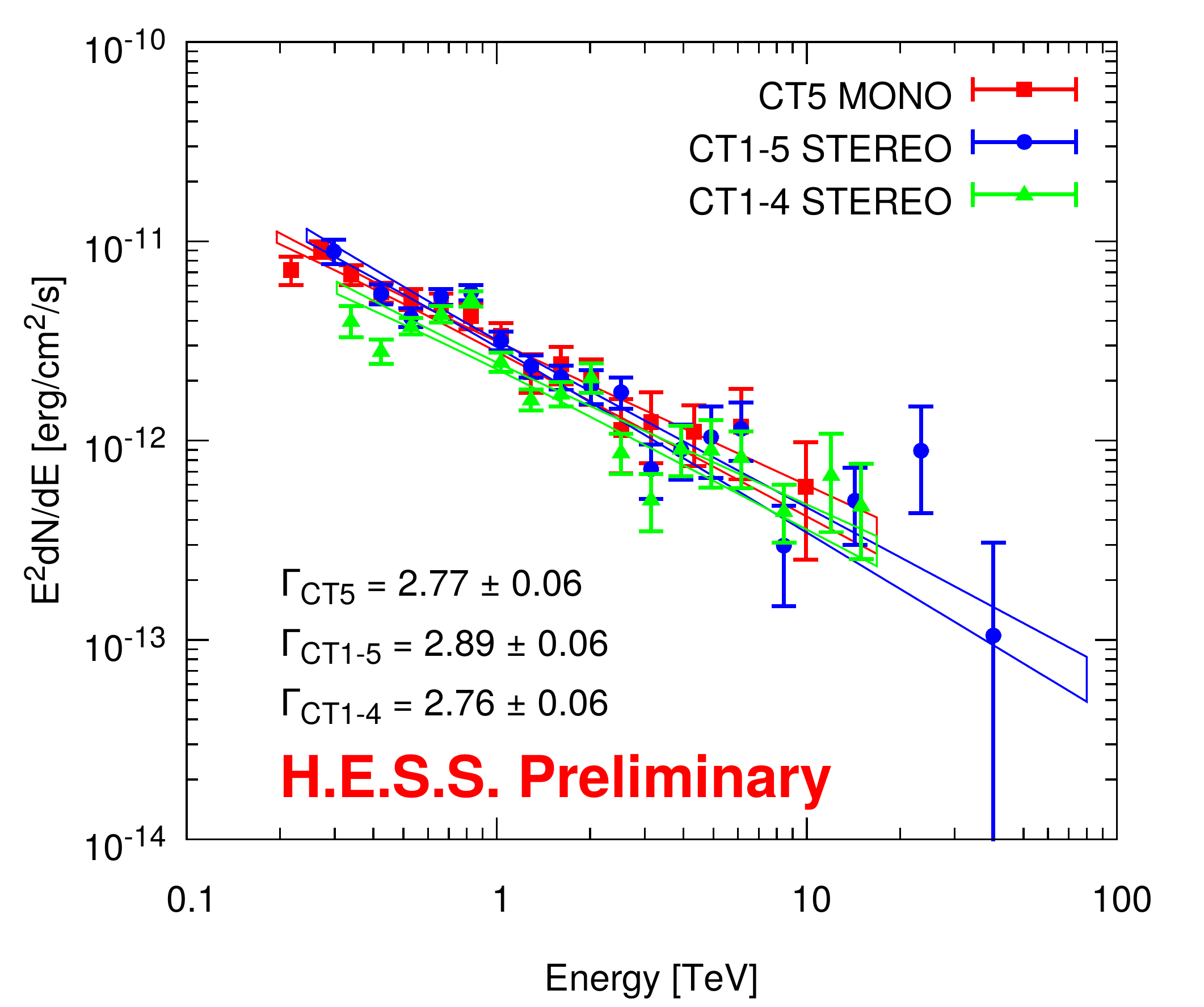}
  \caption{\textbf{Left panel: }Significance distribution around the position of \psrls. \textbf{Right panel:} Spectra obtained from the three different analysis chains for the 2014 data sets.}
\label{fig:skymap_spec}
\end{figure}

In Figure~\ref{fig:skymap_spec} the significance distribution obtained from CT5 MONO data and the spectra obtained from all three analysis configurations are shown. The sky map shows a clear detection of a source that is compatible with a point-like source given the angular resolution described by the point-spread function shown in the inset. The position of the excess is compatible with the position of the \psrls system. North of this source also the extended emission, detected at a significance level of $\sim10\sigma$, from the pulsar wind nebula HESS J1303-631 is visible. The presence of this source was taken into account in the generation of the results.

The spectrum is well described by a power law function with indices ranging from $\Gamma=2.76\pm0.06$ for the CT1--4 STEREO analysis, to $\Gamma = 2.89 \pm 0.06$ found for the CT1--5 STEREO data. The values
are compatible with each other when taking into account the statistical and systematic uncertainties in the reconstruction. The normalization at 1 TeV was reconstructed at values of $\left(1.9\pm0.1\right)\times10^{-12}$ ph/cm$^{2}$/s/TeV for the CT5 MONO and CT1--5 STEREO analysis. The  differential flux at 1 TeV was instead slightly lower in the CT1--4 STEREO analysis with a value of $\left(1.48\pm0.06\right)\times10^{-12}$ ph/cm$^{2}$/s/TeV. This discrepancy is however mitigated when systematic effects, which are on the order of 20--\SI{25}{\percent}, are taken into account.

\section{Phase-folded analysis}

\begin{figure}
\centering
\includegraphics[width=0.9\textwidth]{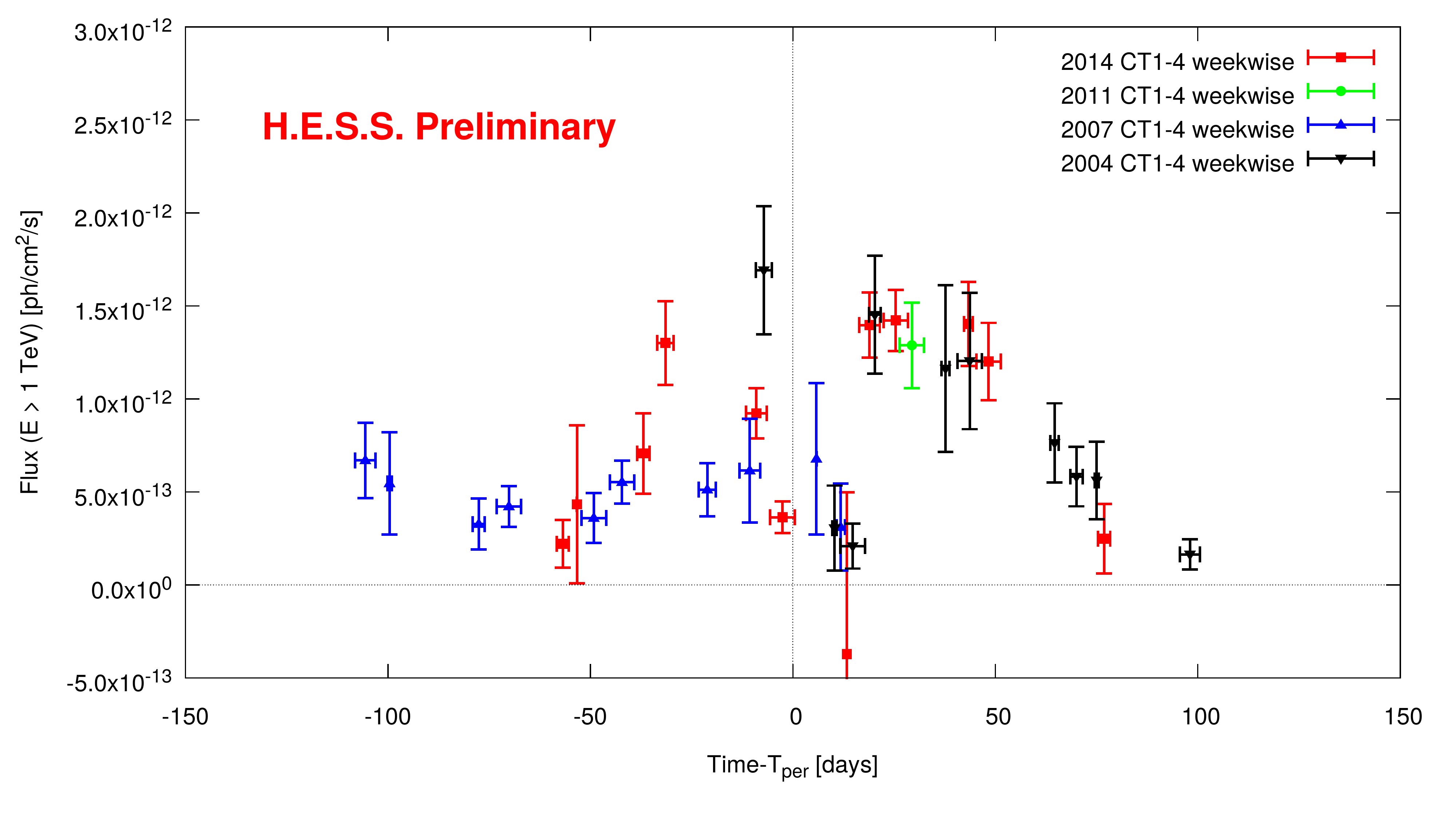}
\caption{Weekly light curve of the CT1--4 STEREO data. Red squares indicate the 2014 data, green circles the 2011 data, blue upward triangles the 2007 data and black downward triangles the 2004 the data.}
\label{fig:weeklyLC}
\end{figure}

\begin{figure}
\centering
\includegraphics[width=0.9\textwidth]{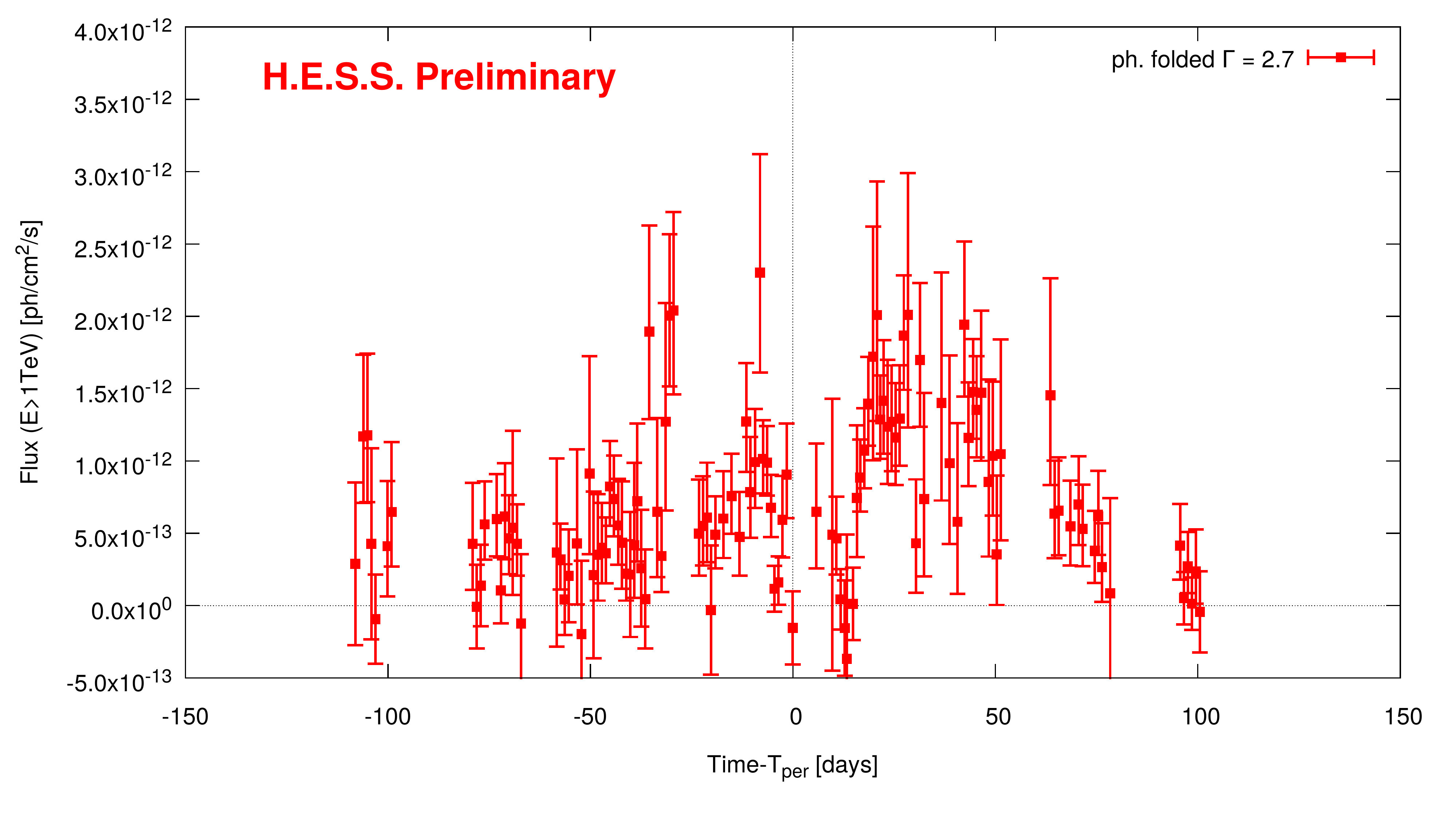}
\caption{Daily phase-folded light curve making use of all CT1--4 STEREO data available.}
\label{fig:phfoldedLC}
\end{figure}

All the CT1--4 STEREO data were re-analysed separately for each year using the latest software version and binned in weekly bins as shown in Figure~\ref{fig:weeklyLC}. The data from the 2014 campaign cover several of the  sampling gaps in the light curve obtained from previous campaigns. Since none of the overlapping flux points deviate from each other significantly, we assume that the same physical processes are responsible for the emission seen in every cycle, and producing a similar (repetitive) gamma-ray output. For the sake of increasing the statistics available at every orbital phase, we stack the data collected in 2004, 2007, 2011 and 2014. For this purpose, each run was analysed individually and the high-level results like the number of excess events or the respective live times were grouped in daily bins, merging data in bins from different years that had the same timely distance from the time of periastron. This phase-folded light curve was produced assuming an average photon index $\Gamma=2.7$. However, the choice of the value of the assumed index has a negligible effect on the flux value in each bin. The outcome is reported in Figure~\ref{fig:phfoldedLC}.

\section{Discussion}

Several interesting features can be noticed in the light curve. Especially in the less noisy weekly light curve is can be seen that there are several parts of the orbit during which the detected emission is at high levels. In particular at the time interval of the GeV flare the source remains in high state. At the time around the periastron passage, we confirm that the flux from this source is in a local minimum. The phase region before the periastron passage is the one that presents the most unexpected features. As already described in \cite{2016arXiv161003264B} and confirmed by this new reanalysis, there is a brightening of the source 30 days before the periastron passage, i.e. before the nominal time of the first disk crossing. Here the flux is $\sim2.5$ times higher than at the moment of the first disk crossing (situated at $t_{per}-\SI{15}{\day}$). The \si{\tev} emission increases approximately ten days before the time of the periastron passage.


\section{Conclusions}

In these proceedings the latest results from the analysis of the gamma-ray binary \psrls are presented. We have shown here the results of the analysis of the 2014 data using all three available telescope and analysis configurations, and for the first time a complete phase-folded lightcurve in daily bins after a reanalysis of the entire available dataset on this source.

The major results are the confirmation of the enhanced emission in correspondence with the \si{\gev} flare, already pointed out in the previous proceedings on this source \cite{2015arXiv150903090R,2016arXiv161003264B}. And more interestingly, we also find a high flux approximately 30 days before the periastron. Such a high flux was not expected beforehand. To investigate this point, observations are scheduled to obtain data in this phase interval close to the next periastron passage, which will take place in September 2017.

\section*{Acknowledgements}
The support of the Namibian authorities and of the University of Namibia in facilitating the construction and operation of H.E.S.S. is gratefully acknowledged, as is the support by the German Ministry for Education and Research (BMBF), the Max Planck Society, the German Research Foundation (DFG), the Alexander von Humboldt Foundation, the Deutsche Forschungsgemeinschaft, the French Ministry for Research, the CNRS-IN2P3 and the Astroparticle Interdisciplinary Programme of the CNRS, the U.K. Science and Technology Facilities Council (STFC), the IPNP of the Charles University, the Czech Science Foundation, the Polish National Science Centre, the South African Department of Science and Technology and National Research Foundation, the University of Namibia, the National Commission on Research, Science \& Technology of Namibia (NCRST), the Innsbruck University, the Austrian Science Fund (FWF), and the Austrian Federal Ministry for Science, Research and Economy, the University of Adelaide and the Australian Research Council, the Japan Society for the Promotion of Science and by the University of Amsterdam.
We appreciate the excellent work of the technical support staff in Berlin, Durham, Hamburg, Heidelberg, Palaiseau, Paris, Saclay, and in Namibia in the construction and operation of the equipment. This work benefited from services provided by the H.E.S.S. Virtual Organisation, supported by the national resource providers of the EGI Federation.

\end{document}